\documentstyle[graphicx]{mn2e}
\newif\ifAMStwofonts

\title{Modelling galaxy clustering:
Is new physics needed in galaxy formation models?}
\author[Han Seek Kim et al.]
       {Han Seek ~Kim,$^1$\thanks{h.s.kim@durham.ac.uk} C.M.~Baugh,$^1$ S.~Cole,$^1$ C.S.~Frenk,$^1$ A.J.~Benson$^2$ \\
       $^1$Institute for Computational Cosmology, Department of Physics, University of Durham, South Road, Durham DH1 3LE, UK\\
       $^2$Theoretical Astrophysics, Caltech, MC350-17, 1200 E. California Blvd., Pasadena CA 91125, USA}
\date{}

\pagerange{\pageref{firstpage}--\pageref{lastpage}}
\pubyear{2009}

\begin{document}

\maketitle
\title{Luminosity dependence of galaxy clustering}
\label{firstpage}

\begin{abstract}
The clustering amplitude of galaxies depends on their intrinsic luminosity. 
We compare the properties of publicly available galaxy formation models 
with clustering measurements from the 
two-degree field galaxy redshift survey. The model predictions show the 
same qualitative behaviour as the data but fail to match the observations 
at the level of accuracy at which current measurements can be made. We 
demonstrate that this is due to the model producing too many satellite 
galaxies in massive haloes. We implement simple models to describe two 
new processes, satellite-satellite mergers and the tidal dissolution of 
satellites to investigate their impact on the predicted clustering. We find 
that both processes need to be included in order to produce a model 
which matches the observations.  
\end{abstract}

\section{Introduction}

The clustering of galaxies encodes information about 
the values of the cosmological parameters and also 
about the physical processes behind the formation and 
evolution of galaxies. In the cold dark matter (CDM) 
hierarchical structure formation theory, galaxies grow 
inside dark matter haloes (White \& Frenk 1991; Cole 1991). The 
formation of structure in the dark matter is governed 
by gravity and can be modelled accurately using N-body 
simulations (e.g. Springel, Frenk \& White 2006). However, 
the fate of baryonic material is much more complicated as 
it involves a range of often complex and nonlinear physical 
processes. The efficiency of galaxy formation is expected 
to depend on the mass of the host dark matter halo (e.g. 
Eke et~al. 2004; Baugh 2006). Modelling the dependence of 
galaxy clustering on intrinsic properties such as luminosity 
offers a route to establish how such properties depend 
upon the mass of the host halo and hence to improve our 
understanding of galaxy formation. 

Recent advances in astronomical instrumentation have produced a wealth 
of information on galaxy clustering. The enormous volume and number of 
galaxies in the two-degree field Galaxy Redshift Survey (2dFGRS; Colless 
et al. 2001) and the Sloan Digital Sky Survey (SDSS York et al. 2000) 
have made possible accurate measurements of clustering for samples of 
galaxies defined by various intrinsic properties (Norberg et~al. 2001, 2002; 
Zehavi et~al. 2002, 2005; Madgwick et~al. 2003; Li et~al. 2006). 
The variation of clustering strength with luminosity tells us how 
galaxies populate haloes and hence about the physics of galaxy 
formation. Any discrepancy between the observational measurements of 
clustering and theoretical predictions points to the need to improve 
the models, either by refining existing ingredients or adding new ones.

The dependence of galaxy clustering on luminosity has been measured 
accurately in the local universe (Norberg et~al. 2001, 2002, 2009, in 
preparation; 
Zehavi et~al. 2002, 2005; Li et~al. 2006). Over the period spanned 
by these studies, galaxy 
formation models have evolved significantly, particularly in the treatment 
of bright galaxies (see, for example, Benson et~al. 2003). The majority 
of current models invoke some form of heating of the hot gas atmosphere 
to prevent gas cooling in massive haloes, in order to reproduce the 
bright end of the galaxy luminosity function. This has implications for 
the correlation between galaxy luminosity and host dark matter halo mass, 
which has, in turn, an impact on the clustering of galaxies. 

Li et~al. (2006) compared the semi-analytical galaxy formation models of 
Kang et~al. (2005) and Croton et~al. (2006), two early models with AGN 
feedback, against measurements of clustering from the SDSS. Qualitatively, 
the models displayed similar behaviour to the real data, but 
did not match the clustering measurements in detail. For example, 
Li et~al. show that as the luminosity varies the predictions of the 
Croton et~al. model change in clustering amplitude by a similar amount 
to the observations. The brightest galaxies are the most strongly clustered 
in the model. However, the clustering strength displays a minimum around 
$L_*$ before increasing again for fainter galaxies. The luminosity 
dependence in the SDSS data, on the other hand, is monotonic. 
Li et~al. speculated that the models predict too many galaxies in massive 
haloes. They demonstrated that the clustering predictions could be improved, 
but not fully reconciled with the data, by removing satellite galaxies by hand.

In this paper, we extend this comparison to the 2dFGRS clustering measurements 
and test the latest galaxy formation models. By using the blue selected 
2dFGRS, we widen the range of physics tested to include the processes  
which influence recent star formation. We compare models produced by different 
groups which allows us to probe different implementations of the physics. 
We reach similar conclusions to those of Li et~al. and investigate physical 
ways to achieve the required reduction in the number of satellites.  

The structure of this paper is as follows. We briefly introduce the 
three semi-analytic models we discuss in Section~\ref{Semi}. 
In Section~\ref{clustering}, we compare the two point correlation 
function results for the 2dFGRS with the theoretical predictions. 
In Section~\ref{physics}, we explore the mechanisms that drive 
clustering, particularly the galaxy luminosity -- host halo 
mass relation and give a step-by-step illustration of how the number 
of galaxies as a function of halo mass (the Halo Occupation Distribution) 
is connected to the clustering amplitude. We empirically determine 
the HOD which reproduces the observed luminosity dependence of clustering 
in Section~\ref{HOD}. We implement simple models for two new physical 
processes in Section~\ref{implication}, to see if we can modify the existing 
models to match the observed clustering. Finally, in Section~\ref{summary}, 
we give a summary and conclusions.

\section{Galaxy formation models}\label{Semi}

\begin{figure}
\includegraphics[width=8.6cm,bb=30 185 570 700]{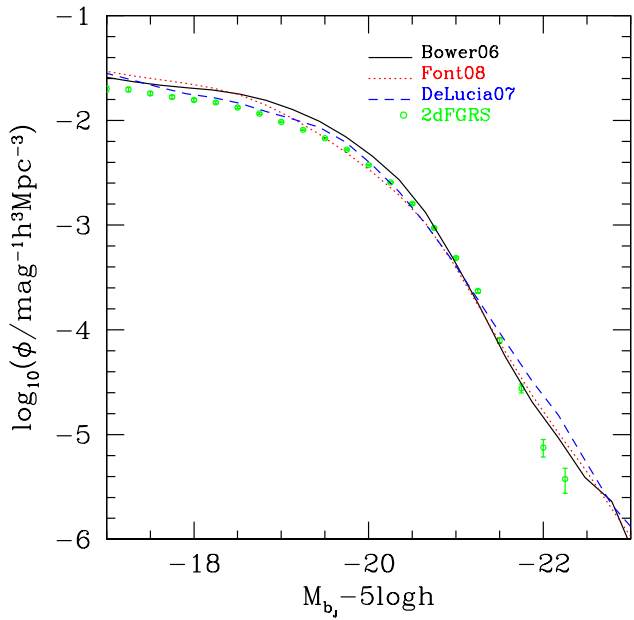}
\caption{
The $b_{\rm J}$-band luminosity function of the Bower et~al. (2006;  
black, solid line), De Lucia \& Blaizot (2007; red, dotted line) and 
Font et~al. (2008) models. The green symbols show the estimate of the 
luminosity function made from the 2dFGRS (from Norberg et~al. 2002).} 
\label{LFALL}
\end{figure}

To make predictions for the clustering of galaxies, we need 
a theoretical tool which can populate large cosmological volumes 
with galaxies. Furthermore, it is essential that we have well 
developed predictions for the properties of the model galaxies,  
in order that we can extract samples which match different 
observational selection criteria. Gas dynamic simulations 
currently struggle to meet both of these requirements. Such calculations 
demand high resolution which limits the accessible computational volume. 
Also, the level of sophistication of the model predictions 
in gas simulations is not always sufficient to make direct contact 
with observational quantities. Semi-analytical models, on the other 
hand, meet both of the above requirements and are therefore well suited 
to clustering studies (for an overview of this approach see Baugh 2006). 

In the first half of this paper we consider predictions for 
galaxy clustering from three semi-analytical models, 
those of Bower et~al. (2006), 
de Lucia \& Blaizot (2007) and Font et~al. (2008). These models 
are publicly available from the Millennium Galaxy Archive\footnote{
http://galaxy-catalogue.dur.ac.uk:8080/Millennium/}. In the second 
part, we consider modifications to the Bower et~al. model. We shall 
also refer to the Bower et~al. and Font et~al. models as the Durham 
models (and as Bower06 and Font08 respectively in figure labels) 
and to the de Lucia \& Blaizot model as the Munich model (and as 
DeLucia07 in plots). 

The three models listed above are set in the context of structure 
formation in a cold dark matter universe as modelled by the Millennium 
Simulation of Springel et~al. (2005). The starting point is the merger 
histories of dark matter haloes, which are extracted from the simulation 
(note both groups have independent algorithms for constructing merger 
histories; see Springel et~al. 2005 and Harker et~al. 2006 for further 
details).  
The models follow a common range of processes which involve the baryonic 
component of the universe: gas cooling, star formation, reheating of cold 
gas by supernovae, chemical evolution of gas reservoirs, heating of 
the hot gas halo by AGN and galaxy mergers. The implementation of these 
processes differs in detail between the models and we refer the reader 
to the original references for a full description.   
Moreover, when setting the model parameters, different emphasis was placed 
on the reproduction of particular observational datasets. 
Here we  simply remark on some key features of the models. 

Bower et~al. (2006) use the model of Malbon et~al. (2007) to describe 
the growth of supermassive black holes through galaxy mergers, and the 
accretion of cold and hot gas. The latter process is the key to matching 
the sharpness of the break in the local optically selected galaxy 
luminosity function. The energy released by the accretion of hot gas 
onto the black hole is assumed to match the luminosity which would have 
been released by gas cooling, thereby 
suppressing the formation of bright galaxies (see Croton et~al. 2006). 
The Font et~al. (2008) model is a development of the Bower et~al. model. 
Firstly, in the Font et~al. model the stellar yield in all modes of star 
formation is twice that adopted in the Bower et~al. model. This shifts 
the locus of the red and blue sequences in the colour magnitude relation 
into better agreement with local data from the Sloan survey 
(see Gonzalez et~al. 2008 for a comparison of the predicted colour 
distributions with SDSS observations). 
Secondly, in the Font et~al. model the stripping of the hot gas from 
newly accreted satellite galaxies is not assumed to be 100\% efficient. 
This is different from the assumption commonly made in semi-analytical 
models and is motivated by the results of recent gas dynamics simulations 
carried out by McCarthy et~al. (2008). This means that in the Font et~al. 
model galaxies can continue to accrete cold gas even after 
they have been subsumed into a more massive halo. 
This results in an improved match to the 
observed colour distribution of satellite galaxies (Gonzalez et~al. 2008). 
Both the Bower et~al. and Font et~al. models give very good matches to 
the stellar mass function over the full redshift range for which observational 
estimates are available.

The De Lucia \& Blaizot (2007) model is a development of the semi-analytical  
models of Springel et~al. (2001), De Lucia et~al. (2004) and Croton et~al. 
(2006). Luminosity and colour dependent clustering were discussed in 
Springel et~al. (2005) and Croton et~al. (2006); the De Lucia \& Blaizot 
model gives similar clustering predictions to those from these earlier 
models. 

The parameters of the models are set to give a reasonable reproduction 
of the present day galaxy luminosity function, as shown by Fig.~\ref{LFALL}. 
In this paper we give ourselves the freedom to adjust the luminosities of 
the model galaxies, whilst maintaining the ranking of galaxy luminosity, 
to force an exact match to the 2dFGRS luminosity 
function measured by Norberg et~al. (2002a). This small adjustment 
allows us to rule out abundance differences as a possible source of 
variations between the clustering predictions of different models. 
We apply the same methodology to the modified versions of the Bower et~al. 
model discussed in the second part of the paper.

\section{Predictions for luminosity dependent clustering}
\label{clustering}

\begin{figure}
\includegraphics[width=8.6cm,bb=30 185 570 700]{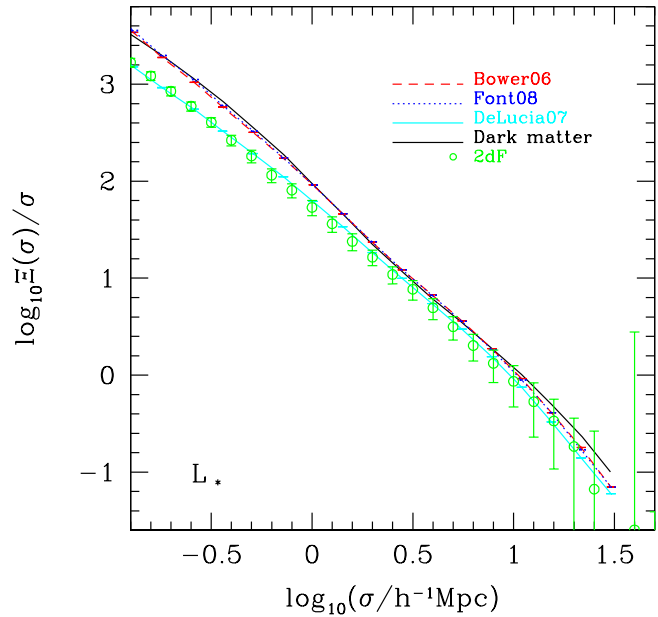}
\caption{\label{2dF2dAll}
The projected correlation function of $L_*$ galaxies 
measured in the 2dFGRS by Norberg et~al. (2009; open symbols). 
The model predictions are shown by different coloured lines, as 
indicated by the key. The projected correlation function of the 
dark matter in the Millennium simulation is shown by the black line. 
}
\end{figure}

\begin{figure*} 
\includegraphics[width=17cm,bb=30 340 565 700]{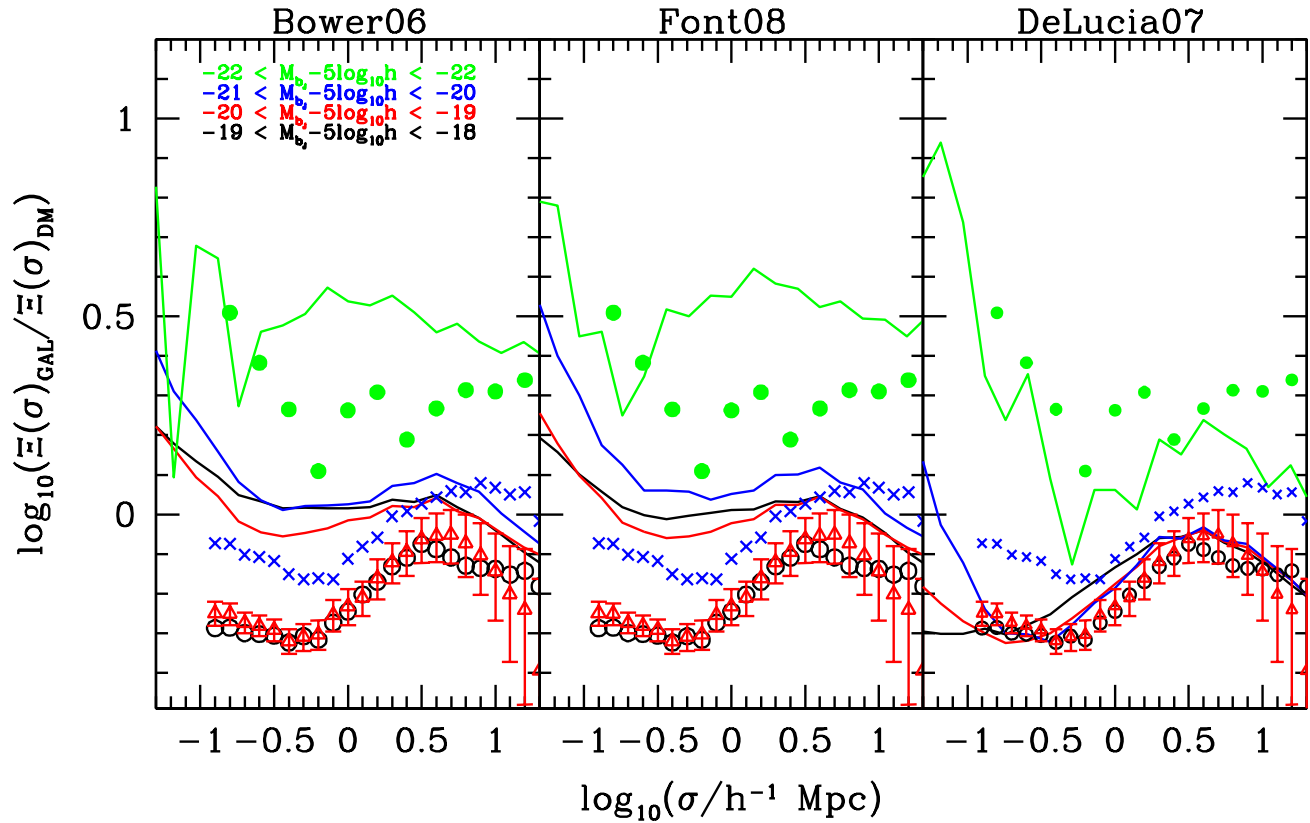}
\caption{\label{DAll}
The projected galaxy correlation functions divided by 
the projected correlation function of the dark matter 
in the Millennium Simulation.  
The symbols show the ratios for the 2dFGRS clustering 
measurements. Different colours show the different luminosity 
bins as indicated by the key. The model predictions are 
shown by the solid lines. Each panel shows the predictions 
for a different model, as indicated by the label. 
}
\end{figure*}

In this section we compare the predictions of the three galaxy 
formation models (Bower et~al. 2006; De Lucia \& Blaizot 2007; 
Font et~al. 2008) with measurements of clustering made from the 
final two-degree field galaxy redshift survey 
(Norberg et~al. 2009b). 
The observational data are presented in the form of the projected 
correlation function, $\Xi(\sigma)/\sigma$. 
This statistic is estimated from the two 
point correlation function binned in pair separation parallel and 
perpendicular to the line of sight, $\xi(\sigma,\pi)$:
\begin{equation}
\frac{\Xi(\sigma)}{\sigma} = \frac{2}{\sigma} \int_{0}^{\infty} 
\xi(\sigma,\pi) {\rm d} \pi.
\label{eq:wrp}
\end{equation}
When redshift is used to infer the radial distance to a galaxy, 
gravitationally induced peculiar motions on top of the Hubble 
flow cause a distortion to the inferred clustering signal. In 
principle, the projected correlation function 
is unaffected by the contribution from peculiar velocities. 
In practice, the integration in Eq.~\ref{eq:wrp} has to be 
truncated at a finite value of $\pi$ as the clustering signal 
on larger scales becomes noisy. Norberg et~al. 
(2009a) show that this truncation has a negligible effect on the 
form of the projected correlation function on scales below $10h^{-1}$Mpc. 

The observational measurements we use in this paper are from 
the final 2dFGRS. Previous results for the luminosity dependence 
of galaxy clustering were presented by Norberg et~al. 
(2001,2002). These papers analysed an intermediate version of 
the 2dFGRS which consisted of around 160\,000 unique, high 
quality galaxy redshifts. In the final version of the dataset 
used by Norberg et~al. (2009) there are more than 220\,000 galaxy 
redshifts. The solid angle of high spectroscopic completeness 
regions has also increased, by a larger factor than the change 
in the total number of redshifts. Hence a more accurate 
measurement of the clustering in different volume limited samples 
is now possible. The estimation of errors on the clustering 
measured for the different samples has also been revisited 
(Norberg et~al. 2008). An internal estimate of the error is made 
using the bootstrap resampling technique. This has the advantage 
over the mock catalogues used previously that the change in 
clustering strength with luminosity is taken into account. The 
2dFGRS is selected in the blue $b_{\rm J}$ band. This is more 
sensitive to recent episodes of star formation in galaxies than 
the red $r$ band selection used in the SDSS. 
 
We first examine the clustering of $L_*$ galaxies. 
Fig.~\ref{2dF2dAll} compares the model predictions and 
the 2dFGRS measurement for the projected 
correlation function of $L_{*}$ galaxies, along with 
the projected correlation function of the dark matter 
in the Millennium Simulation. 
On large scales, $\sigma > 3 h^{-1}$Mpc, the models 
have a similar shape to the observations, but different amplitudes. 
The Durham models (Bower et~al. and Font et~al.) have a higher clustering 
amplitude than the data, but are similar to the dark matter. On small 
scales, $\sigma \le 1 h^{-1}$Mpc, the Durham models are significantly above 
the 2dFGRS measurement. The De Lucia \& Blaizot prediction is a 
remarkably good match to the $L_*$ clustering data over the full range of 
scales plotted. As we will see in the next section, the clustering predictions 
can be broken down into contributions from the most massive 
galaxy in each halo, referred to as the central galaxy, and satellite 
galaxies. The form of the projected correlation function on small scales 
is driven by the number of satellites in massive haloes. One interpretation 
of the comparison in Fig.~\ref{2dF2dAll} is that massive haloes in 
the Durham models contain more $L_*$ satellites relative to low mass haloes 
than in the Munich model. This would also 
account for the small difference 
between the predicted clustering amplitudes on large scales.

Over a range of just over two decades in projected pair separation, 
Fig.~\ref{2dF2dAll} shows that the clustering amplitude changes 
by four and a half orders of magnitude. In order to see more 
clearly the changes in the clustering amplitude with varying galaxy 
luminosity, in Fig.~\ref{DAll} we divide the galaxy correlation 
functions by the dark matter correlation function. If the Millennium 
Simulation dark matter was indeed a match to the real Universe, then 
the ratio plotted in Fig.~\ref{DAll} would be the logarithm of the 
square of the bias, albeit quantified in terms of projected 
clustering. The departure of this ratio from a constant value 
would then indicate the presence of a scale-dependent bias. 
However, it is of course possible that 
the Millennium Simulation is not quite representative of reality, with recent 
studies suggesting a lower value of the fluctuation amplitude 
$\sigma_{8}$ (Sanchez et~al. 2009; Li \& White 2009). Nevertheless,  
the Millennium dark matter serves as a useful benchmark, even if these 
caveats limit the interpretation of the ratio. 

The Durham models overpredict the clustering displayed by the 
brightest 2dFGRS galaxy sample, whereas the Munich model predicts 
weaker clustering for this sample. The Durham models overpredict 
the clustering displayed by the remaining, fainter luminosity samples. 
The Munich model comes closest to reproducing the trends seen in the 
data. As we commented earlier, the Munich model gives a very good match 
to the clustering measured for the $L_{*}$ sample. The amplitude 
of clustering in the magnitude bins either side of the $L_*$ sample 
hardly changes in the De Lucia \& Blaizot model. The largest disagreement 
between that model and the 2dFGRS measurements occurs 
for the $-21 < M_{b_{\rm J}} - 5 \log h < -20$ sample. 

The correlation function ratios plotted in Fig.~\ref{DAll} show strong 
scale dependence. On the largest scales plotted, this could indicate 
that the clustering of dark matter in the Millennium cosmology is 
not the same as in the real Universe, as we remarked upon above. 
However, the 2dFGRS measurements become noisy on the scales on which 
one would expect the bias to approach a constant value (e.g. Coles 1993). 
On small scales there is a range of shapes and amplitudes, indicating 
a wide variety of satellite fractions in the different galaxy samples. 
Apart from the brightest sample, the Durham models show a higher clustering 
amplitude on small scales than the Munich model and also a higher amplitude 
than the observations. This suggests that there are too many satellite 
galaxies in haloes in the Durham models, a conclusion which we confirm 
in the next section. 

\section{What drives galaxy clustering?}
\label{physics}

\begin{figure}
\includegraphics[width=8.5cm]{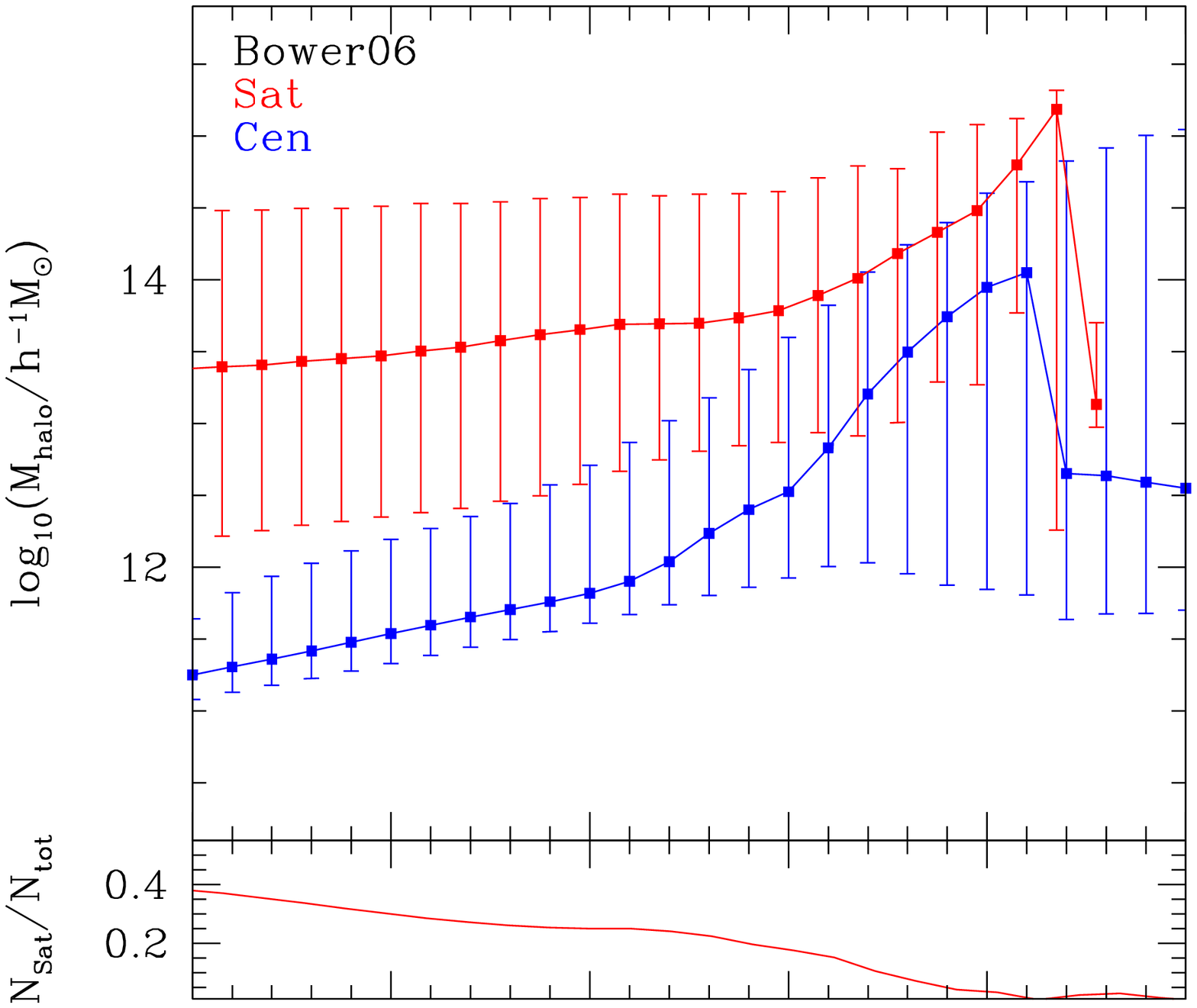}\vspace{-2.05cm}
\vspace{-2.05cm}
\includegraphics[width=8.5cm]{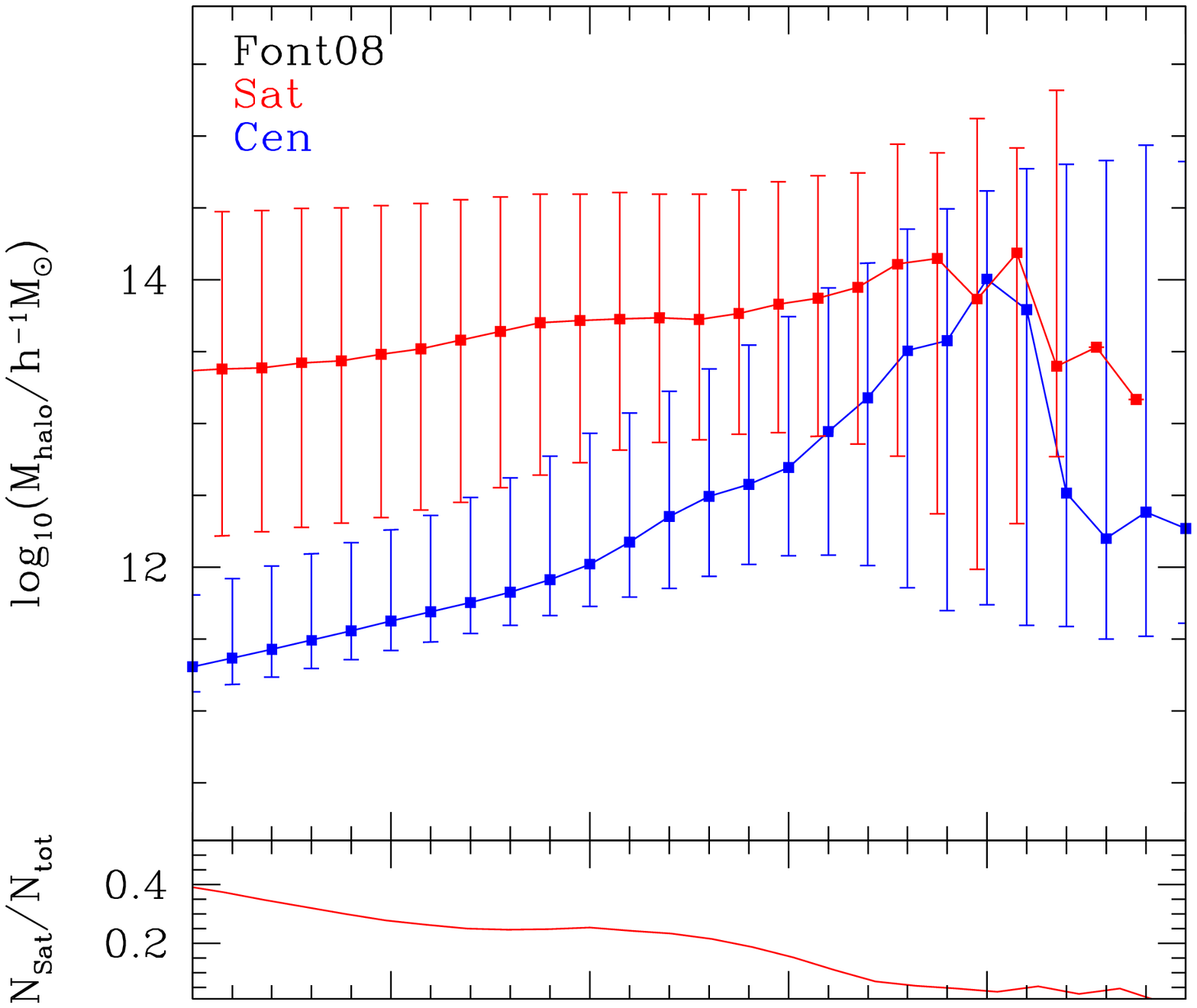}
\includegraphics[width=8.5cm]{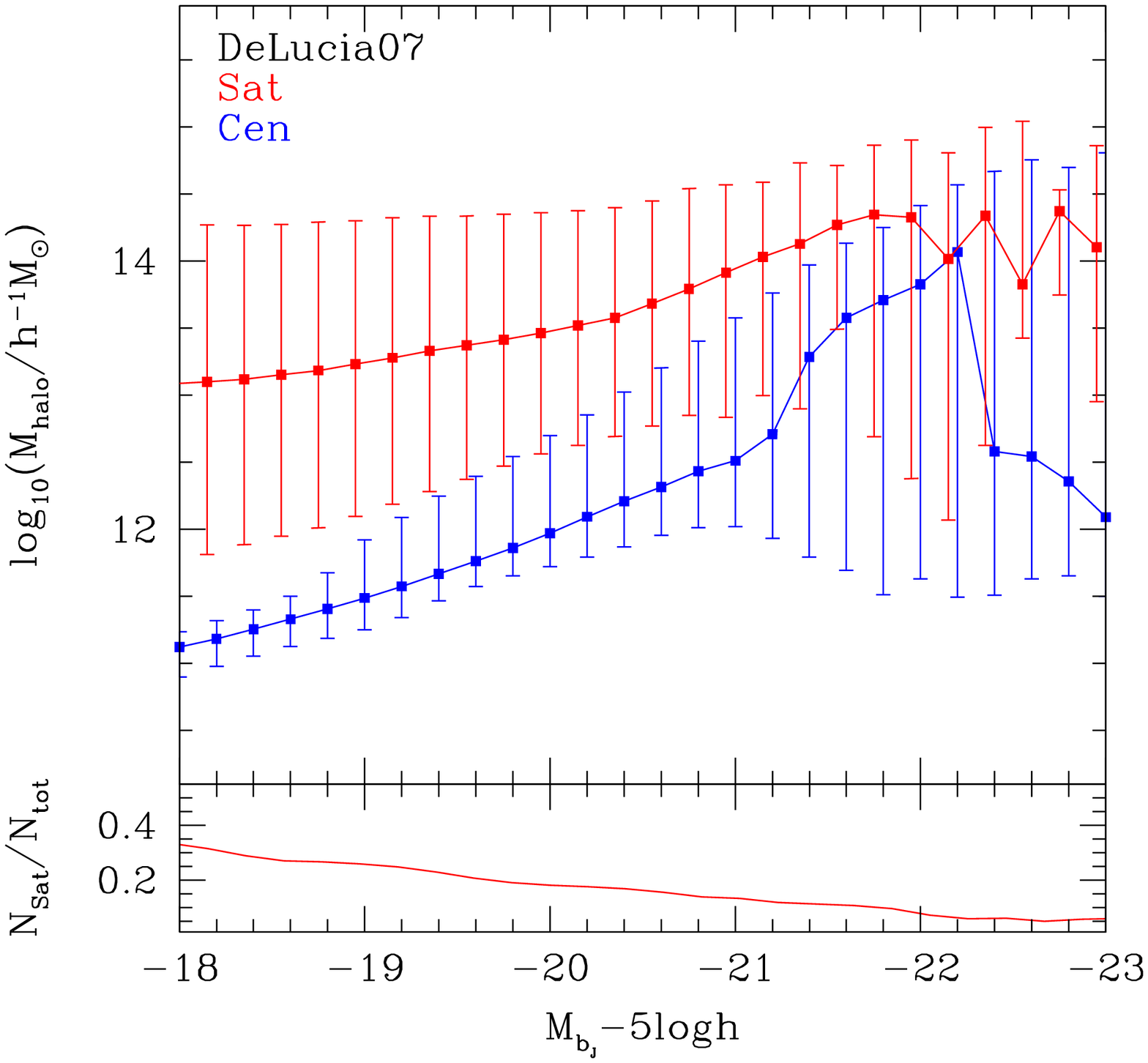}
\caption{\label{MMBJB}
The host halo mass for galaxies as a function of luminosity.
The main window in each panel shows the predictions for a 
different galaxy formation model, with Bower et~al. shown 
in the top panel, Font et~al. in the middle panel and de 
Lucia \& Blaizot in the lower panel. The median mass and 
10-90 percentile ranges are shown separately for central (blue)
and satellite (red) galaxies. The small window in each panel 
shows the fraction of galaxies that are satellites as a function 
of magnitude. 
}
\end{figure}

\begin{figure*} 
\includegraphics[width=17cm,bb=35 185 570 700]{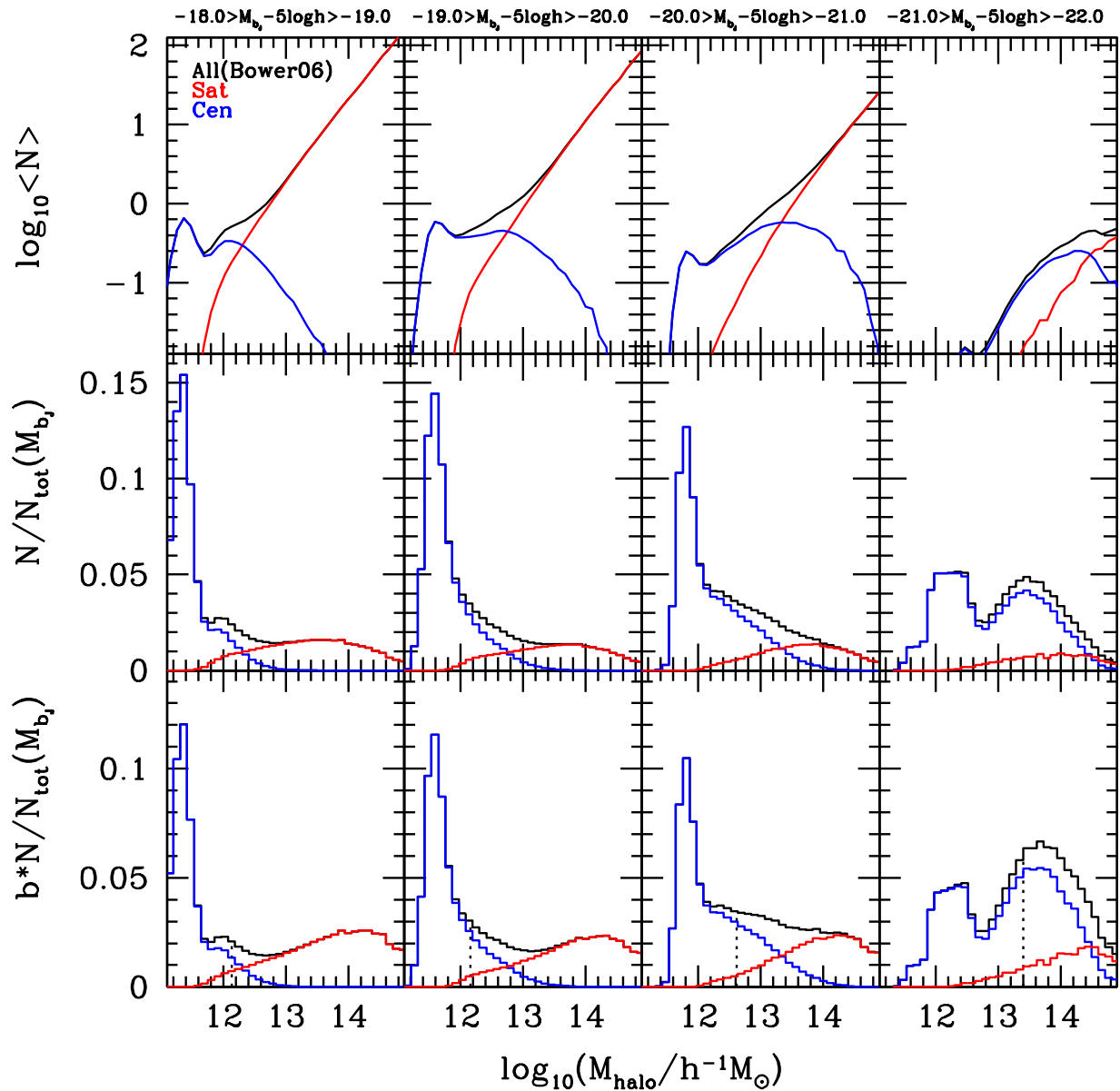}
\caption{\label{BBN} 
The steps connecting the number of galaxies per halo to the strength of 
galaxy clustering in the Bower et~al model. Each column corresponds 
to a different galaxy sample, as indicated by the label. 
The blue curves show the contribution 
from central galaxies, the red curves show satellite galaxies 
and the black curves show centrals plus satellites. The top row 
shows the galaxy halo occupation distribution. The middle row 
shows this HOD multiplied by the dark matter halo mass function and 
normalized by the total number of galaxies in the luminosity bin. 
The bottom row shows the HOD multiplied by the halo mass function 
and the halo bias, again normalized by the total number of galaxies 
in the luminosity bin. In this case the area under the black curve is 
the effective bias of the sample. 
The dotted line in the lower panels shows the mass which 
divides the area under the curve in half.
}
\end{figure*}

\begin{figure*} 
\includegraphics[width=18cm,bb=25 250 570 700]{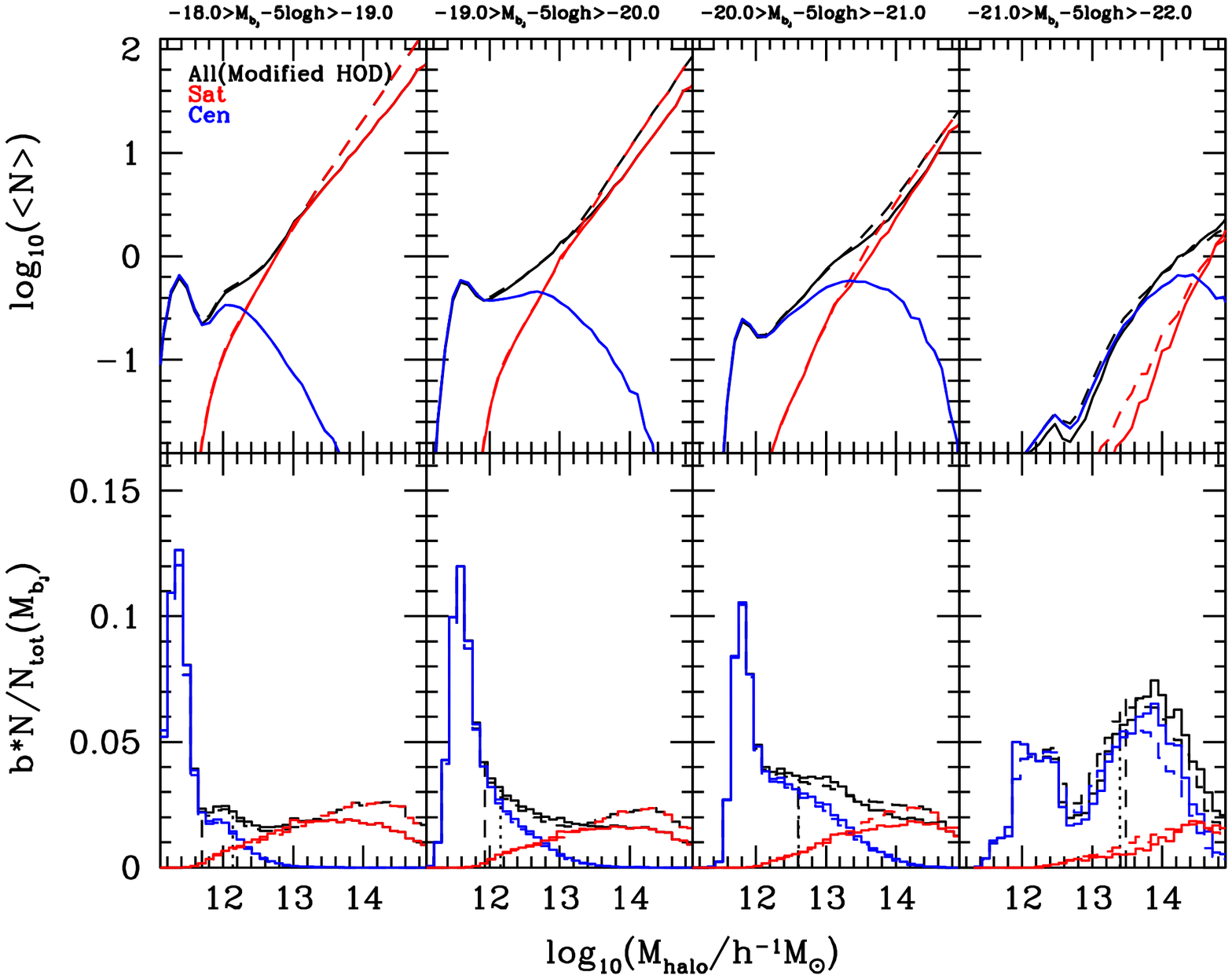}
\caption{\label{BMN}
{\it Top row}: A comparison of the modified HOD (solid lines) 
in which the slope of the satellite HOD has been adjusted to 
match the 2dFGRS clustering measurements with the original 
HOD of the Bower et~al. model (dashed lines). 
{\it Bottom row}: The contribution to the effective bias as a 
function of halo mass. The quantity plotted is the modified HOD 
weighted by the halo mass function and the halo bias parameter; 
the area under the black curve gives the effective bias parameter. 
Each column corresponds to a different luminosity bin as shown 
by the label. The blue curves show the contribution of central 
galaxies, satellites are shown in red and the total is shown 
in black. The vertical lines mark the halo mass which divides 
the contribution to the effective bias integral into two. 
The dotted lines show this mass for the original Bower et~al. 
model and the dashed lines for the modified HOD. 
}
\end{figure*}

In this section, we look at the clustering predictions in more detail to 
identify which galaxies determine the shape and amplitude of the correlation 
function. This will allow us to identify which model galaxies are responsible 
for the disagreement found with observational measurements in the previous 
section, and hence will motivate approaches to altering the model 
predictions for these objects. 

The clustering of dark matter haloes depends on their mass. Haloes which are 
more massive than the characteristic mass scale at a particular redshift 
(roughly the location of the break in the halo mass function) will be much 
more strongly clustered than the overall dark matter (Cole \& Kaiser 1989; 
Mo \& White 1996). We start by plotting 
the relation between galaxy luminosity and the mass of the host dark matter 
halo in Fig.~\ref{MMBJB}. The main panel in each plot shows the median host 
halo mass and 10-90 percentile range of the distribution as a function of 
luminosity, for satellite and central galaxies separately. The sub-panel 
shows the fraction of galaxies that are satellites at each magnitude. 

Overall, the host halo mass -- galaxy luminosity relations for the different 
models share the same qualitative behaviour. There is a trend of increasing 
host mass with increasing central galaxy luminosity which steepens around 
$M_{b_{\rm J}} - 5 \log h \approx -21$. A magnitude brighter than this, the 
median host halo mass drops in each case. The scatter in host mass 
is small at the faintest luminosities plotted (around a factor of 2 in the 
de Lucia \& Blaizot model), and increases with luminosity. For the brightest 
galaxies shown in Fig.~\ref{MMBJB}, the 10-90 percentile range covers more 
than 2 orders of magnitude in halo mass. The median host mass of satellite 
galaxies does not increase with luminosity as quickly as it does for the 
centrals (an order of magnitude increase in host mass over the magnitude 
range $-18 > M_{b_{\rm J}} - 5 \log_{10} h > -22$, compared with two orders of 
magnitude for the centrals). The 10-90 percentile range is very broad for 
faint and intermediate luminosity satellites ( $\sim 2$ orders of magnitude) 
and shrinks only for the brightest satellites.

The quantitative differences between the models in Fig.~\ref{MMBJB} explain 
the differences in the predictions for luminosity dependent clustering evident 
in Fig.~\ref{DAll}. Firstly, the median host mass relations for the central 
and satellite galaxies in the de Lucia 
\& Blaizot model are lower than those in the Bower et~al. and Font et~al. 
models. This means that the overall amplitude of clustering is lower in the 
de Lucia \& Blaizot model as seen in Fig.~\ref{DAll}. Secondly, the scatter in 
the mass -- luminosity relation for centrals is substantially smaller in 
the Munich model than it is in the Durham models, particularly for fainter 
galaxies. This means that the halo mass -- luminosity relation is better 
defined in the Munich model compared with the Durham models, which explains 
the somewhat stronger trend of luminosity dependent clustering displayed 
in the Munich model. 

The difference in the width of the distribution for the central galaxies 
could be driven by the choice of time over which gas is allowed 
to cool in a halo. 
In the Munich model, gas is allowed to cool over a dynamical time. In the 
Durham models, the cooling time depends upon the merger history of the 
individual trees. For haloes of a given mass, there will therefore be a range 
of cooling times in the Durham models, but a fixed cooling 
time in the Munich model. 

The subpanels in each part of Fig.~\ref{MMBJB} show the fraction of satellite 
galaxies as a function of luminosity. For all the models, the fraction 
declines to brighter magnitudes. Due to the wide range of halo masses 
occupied by satellites, and the strong dependence of bias or clustering 
strength on halo mass, it is possible for satellites to make an important 
contribution to the overall clustering signal, even if they are outnumbered 
by centrals. We investigate this point in more detail next in this section. 
than in the Munich models. Furthermore, there is a plateau in the satellite 
The fraction of satellites in the Durham models is somewhat higher 
fraction for intermediate luminosities in the Durham model which is not 
present in the Munich models. This suggests that we should focus on reducing 
the number of satellite galaxies in order to improve the Durham model 
predictions for luminosity dependent clustering. 

An alternative way to present the information contained in the host mass -- 
luminosity plot is the halo occupation distribution (HOD; for a review see 
Cooray \& Sheth 2002). The HOD gives the mean number of galaxies as a 
function of dark matter halo mass, divided into the contribution from central 
and satellite galaxies. The HOD has the advantage over the host halo 
mass -- galaxy luminosity plot that it can be more directly 
related to galaxy clustering (e.g. Benson et~al. 2000; Berlind et~al. 2003). 

The HOD for the Bower et~al. model is shown in the top row of 
Fig.~\ref{BBN}, in which each column shows the HOD for galaxies 
in a different bin in absolute magnitude. The bins are one magnitude 
wide, whereas in the majority of cases 
in the literature, cumulative bins are used. The generic form adopted for 
the HOD is a step function for central galaxies, which makes the transition 
from 0 to 1 galaxies per halo at some halo mass threshold, which is determined 
by the galaxy selection (e.g. Zehavi et al. 2002). More gradual forms for the 
transition from 0 to 1 galaxy per halo have been discussed 
(Zheng et~al. 2005). The HOD for satellites is assumed to be a power-law 
with slope $\alpha$; the mean number of satellites per halo reaches unity 
at a somewhat higher halo mass than that at which the mean number of central 
galaxies first approaches unity. The satellite galaxy HOD for 
the Bower et~al. model agrees with the standard HOD paradigm. The central 
galaxy HOD, on the other hand, has a richer structure. The downturn seen 
at high masses is due to the adoption of a differential, finite width 
magnitude bin. With increasing halo mass, the central galaxies eventually 
become too bright to be included in a particular magnitude bin. For all 
the luminosity bins plotted, the HOD of central galaxies does not reach 
unity, in contradiction to one of the primary assumptions in HOD modelling. 
The central HOD rises to a peak just below unity, before showing a dip with 
increasing halo mass. This feature is due to AGN heating  which suppresses 
gas cooling above $M \sim 10^{12} h^{-1}M_{\odot}$ at the present 
day in this model. This spike has a similar appearance in the  
Font et~al. model, even though the ``switch-on'' of AGN heating 
feedback is handled in a more gradual way in this case. 

The HOD does not tell us the full story about galaxy clustering, but  
is only the first step. The next relevenat consideration is 
the abundance of dark matter haloes. The number density of haloes declines 
exponentially with increasing mass beyond the characteristic mass (see for 
example Jenkins et~al. 2001). 
The HOD 
weighted by the halo mass function is shown in the second row of 
Fig.~\ref{BBN}. Note that 
we have now switched to a linear scale on the y-axis. 
The contribution of 
satellite galaxies is now much less important than the impression gained 
from the HOD plot. Next, in the bottom row of Fig.~\ref{BBN} we plot, 
as a function of halo mass, the HOD multiplied 
by the halo mass function and the bias factor.
Again, a linear scale is used for the y-axis. The area under the black 
curve in this case gives the effective bias of the galaxy sample. 
The satellites make 
a larger contribution to the effective bias than they do to the number 
density. This is because the satellites are preferentially found in high 
mass haloes which have large bias factors.

\section{An empirical solution to the problem of 
luminosity dependent clustering}
\label{HOD}

\begin{figure} 
\includegraphics[width=8.6cm,bb=30 185 570 700]{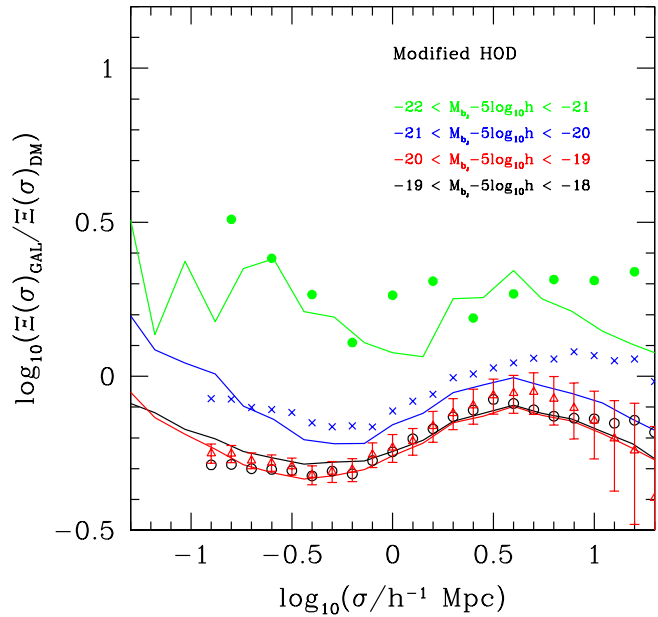}
\caption{\label{DFMHOD}
The clustering of galaxies after modifying the HOD of the Bower et~al. 
model (lines) compared to the 2dFGRS data 
(points). We plot the projected correlation function divided by an analytic 
estimate of the nonlinear projected correlation function of the dark matter 
in the Millennium simulation cosmology. Different colours show the results 
for different luminosity bins as indicated by the key.  
}
\end{figure}

In this section we find an empirical solution to the 
problem of matching the observed luminosity dependence 
of clustering. We do this by changing the HOD of the 
Bower et~al. model by hand. We could equally well 
have chosen to use the Font et~al. model 
and would have reached similar conclusions. 
We saw in the previous section that the HOD for central 
galaxies has a complicated shape which is not well described 
by the standard HOD parametrizations. 
This is, in part, due to the physics invoked in the models 
and to the use of differential 
rather than cumulative luminosity bins. The satellite 
galaxy HOD, on the other hand, has a more straightforward 
power law form, $N_{\rm sat} \propto M_{\rm halo}^{\alpha}$, 
where $M_{\rm halo}$ is the host halo mass. Moreover, we 
saw in the previous section that the Durham models have more 
satellite galaxies than the Munich model and that this could 
be the reason behind their poorer match to the observed clustering. 
Here, we establish how the satellite HOD must be changed in 
order to match the 2dFGRS results better. This will help guide 
an investigation into changing the physics of the galaxy 
formation model which is carried out in the next section.

The satellite HOD for the Bower et~al. model plotted in Fig.~\ref{BBN} 
has a power law form with slope $\alpha \sim 1$ in each of the 
luminosity bins. We note that the same slope is generally found 
for other galaxy selections, such as luminous red galaxies 
(Almeida et~al. 2008; Wake et~al. 2008). 

The starting point to make a realization of galaxy clustering is the 
DHalo\footnote{http://galaxy-catalogue.dur.ac.uk:8080/Millennium/} 
catalogue of dark matter halo masses and positions constructed from 
the Millennium simulation (Harker et~al. 2006). This is the halo 
catalogue used in the {\tt GALFORM} model and is somewhat different 
from the list of haloes generated by the friends-of-friends 
group finding algorithm. The DHalo catalogue is constructed with reference 
to the merger histories of the dark matter haloes. In the case of a 
friends-of-friends merger history, it is possible, occasionally, for the 
mass of a halo to decrease with increasing time. This happens, for example,
 when two haloes are either extremely close or overlap to some extent at 
one timestep, but move apart 
and are identified as separate haloes at a subsequent output time. 
The DHalo algorithm ``looks ahead'' to check if haloes 
merged by the group finder at one output time stay merged at the next 
two outputs. 

Keeping the same mass at which the mean number of satellites per halo 
reaches unity as predicted by the fiducial Bower et~al. model, 
we allow the slope of the satellite HOD to vary for each magnitude 
bin in order to obtain a better match to the 2dFGRS clustering data. 
The number of galaxies 
as a function of halo mass is assumed to have a Poisson distribution 
for $N>1$. For halo masses for which the HOD predicts $N<1$, a fraction of 
haloes is populated with a satellite galaxy at random: i.e. if the 
random number chosen from a uniform distribution between zero and one, 
$x<N$, then the halo is assigned a satellite, otherwise 
it has no satellite. We have tested that this procedure can 
reproduce the clustering in the Bower et~al. model when the Bower et~al. 
HOD is used.  

The modified HOD derived as described above is shown by the solid lines 
in Fig.~\ref{BMN}. The HOD of the original Bower et~al. model is shown 
by the dashed lines in this plot. In the three faintest luminosity bins, 
the slope of the modified satellite HOD is shallower than the 
original i.e. $\alpha < 1$, corresponding to a reduction in the number 
of satellites in massive haloes. The change in slope is largest in the 
faintest bin. In the brightest luminosity bin, the trend is reversed and 
there are slightly more satellites in massive haloes in the modified HOD. 
By reducing the number of satellites in high mass haloes, two effects are 
generated in the correlation function. The effective asymptotic bias of the 
sample is reduced, due to a smaller two-halo clustering term. Also, 
the one-halo term is suppressed, reducing clustering on small scales, 
as there are fewer pairs 
of galaxies within massive haloes. By contrast with the modified HOD, 
as we remarked upon above, the HOD of the Bower et~al. model exhibits the 
same value of the slope of the satellites in each luminosity bin. 

Fig.~\ref{DFMHOD} shows that the trend of clustering strength with luminosity 
displayed by the modified HOD matches that of the 2dFGRS data. Furthermore, 
the improved level of agreement is seen on both large and small scales. 
The matching of the asymptotic bias on large scales and the shape 
of the correlation function on small scales is convincing evidence in 
support of the modified HOD having 
the correct number of satellite galaxies in haloes of different masses. 
The challenge now is to see if the semi-analytical model can reproduce 
the form of the modified HOD, either by further exploration of the model 
parameter space or by adding new physical processes.

\section{Implications for satellite galaxies 
in galaxy formation models}\label{implication}

In the previous section we demonstrated that the clustering properties  
of the Bower et~al. model can be significantly improved if the number of 
satellite galaxies in massive haloes is reduced. This was achieved by changing 
the HOD of the Bower et~al. model by hand. The 
clustering predictions subsequently changed on all scales (in HOD 
terminology, both the one and two halo contributions  
were changed) to improve the match with the 2dFGRS measurements, which 
can only be achieved by changing the number of satellites. In this section 
we try to reproduce the modified HOD in a physical, rather than 
empirical, way by using the {\tt GALFORM} model. 

The first approach we tried was to run variants of the Bower et~al. model 
in which selected parameters were perturbed from their 
fiducial values. In particular, we varied parameters which we thought would 
have an impact on the relation between galaxy luminosity and host halo mass, 
as plotted in Fig.~\ref{MMBJB}. These included the strength of supernova 
feedback, the degree of suppression of gas cooling in massive haloes due 
to AGN heating and the timescale for galaxy mergers. In the case of each 
of these variant models, we 
rescaled the model galaxy luminosities to agree exactly with the 
observational estimate of the luminosity function from Norberg et~al. 
(2009). The clustering predictions in the variants were different 
to those of the original Bower et~al. model. However, none was able to match the observed 
clustering. Intriguingly, the slope of the satellite HOD was $\alpha 
\approx 1 $ in all of the models, that is none of the parameter 
variations was able to change the slope of the satellite HOD in the way 
suggested by the modified HOD. 

The second approach we tried was to change the timescale for galaxies to 
merge due to dynamical friction. {\tt GALFORM} uses a modified 
version of the timescale given by the dynamical friction formula of 
Chandrasekhar (1943; see eqn. 4.16 of Cole et~al. 2000). 
We experimented with adjusting this timescale 
by allowing an extra scaling based on the ratio of the host halo mass 
to the mass of the satellite, $M_{\rm H}/M_{\rm sat}$. To solve the 
problem of too many satellites we needed to reduce the merger timescale 
for $M_{\rm H}/M_{\rm sat}>1$. Recent numerical studies of satellite 
mergers found that the Chandrasekhar formula needs to be revised but in the 
opposite sense, i.e. with a somewhat longer merger timescale for 
objects with $M_{\rm H}/M_{\rm sat}>1$ (Jiang et~al. 2008, 2009). Hence 
this approach, although viable, was abandoned as requiring an unrealistic  
change to the prescription for calculating the timescale for galaxy mergers. 

In this section, we explore the incorporation of two physical processes 
into the {\tt GALFORM} semi-analytical model: the tidal disruption or 
stripping of mass from satellite galaxies and mergers between satellites. 
The implementations 
presented here are exploratory and are meant to give an indication of 
the likely impact of the new physics on the model predictions. If the 
changes turn out to be promising, the intention is that this should 
motivate future, fully self-consistent revisions to the {\tt GALFORM} 
machinery.

\subsection{The dissolution of satellite galaxies}
\label{ICLeffect}

\begin{figure} 
\includegraphics[width=8.6cm,bb=30 185 570 700]{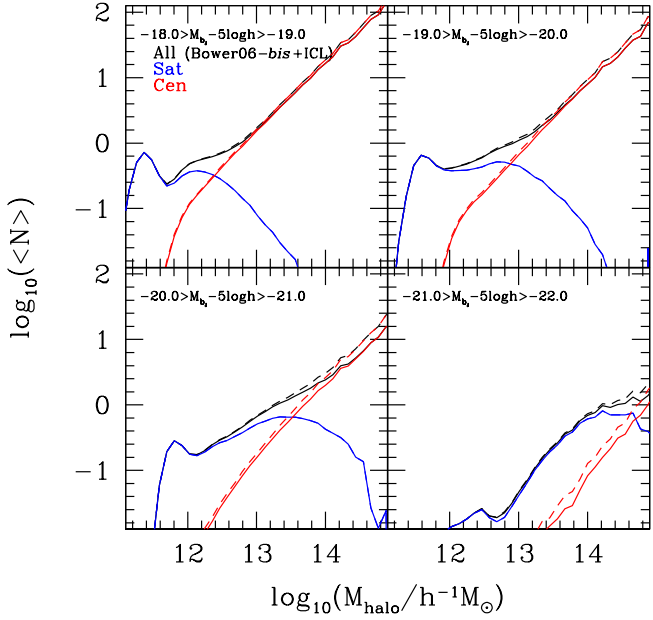}
\caption{\label{HODBICL}
The HOD after applying the satellite disruption model of Eq.~\ref{ICLapply} 
(solid lines). The starting point is the HOD of the Bower06-{\it bis} model 
shown by the dashed lines. Each panel corresponds to a different luminosity 
bin as indicated by the key.    
}
\end{figure}

\begin{figure} 
\includegraphics[width=8.6cm,bb=30 185 570 700]{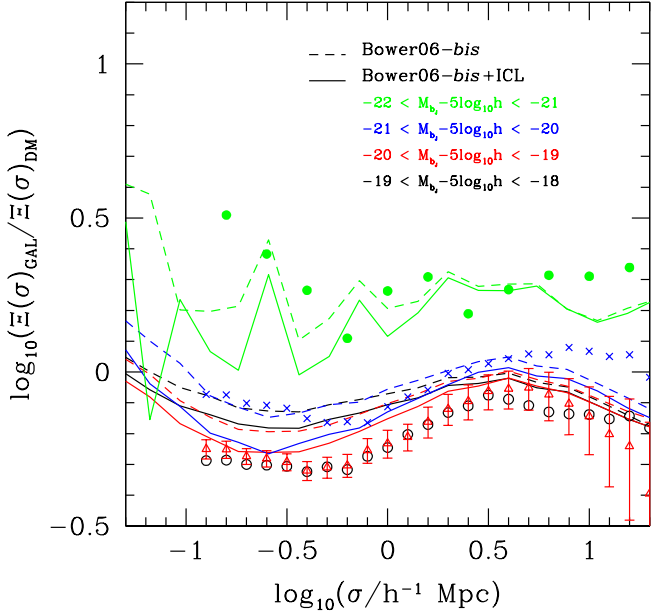}
\caption{\label{DFBICL} 
The projected correlation function for galaxy samples of different 
luminosity divided by the dark matter projected correlation function 
for the Millennium simulation cosmology. The dashed lines show the 
predictions of the Bower et~al. (re-run) model and the solid lines show 
this model after applying the satellite disruption model of Eq.~\ref{ICLapply}.
The symbols show the clustering data measured from the 2dFGRS.  
}
\end{figure}

\begin{figure} 
\includegraphics[width=8.5cm,bb=30 185 580 700]{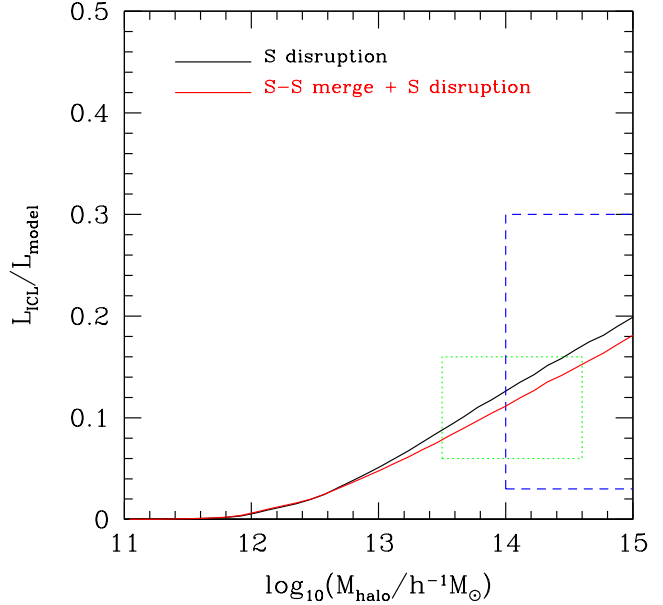}
\caption{\label{LILB}
The intracluster light as a function of halo mass in the satellite disruption 
model. The y-axis shows the fraction of the total cluster light which is 
attached to galaxies. 
The green box shows the observational 
estimate of the intracluster light from Zibetti (2008) and the 
blue box shows the result from Krick \& Bernstein (2007).
The red line shows the intracluster light predicted by the model with 
satellite disruption alone (as discussed in Section 6.1); 
the black line shows a model with satellite-satellite mergers (Section 6.2) 
and disruption of satellites. This hybrid model is discussed in 
Section 6.3. 
}
\end{figure}

Galaxy clusters contain a diffuse background of light, the intracluster 
light (ICL), which is not associated with any particular galaxy (e.g. Welch \& 
Sastry 1971). The ICL is thought to result from the disruption of small 
galaxies and the stripping of stars from larger ones. The measurement 
of the intracluster light is challenging. Current estimates put the ICL 
in the range of 5-30\% of the total cluster light (Zibetti et~al. 
2005; Krick \& Bernstein 2007; Zibetti 2008).

A number of physical processes could be responsible for the removal of stars 
from satellite galaxies e.g. tides produced by the cluster potential and successive high speed fly-by encounters between cluster members 
(Richstone 1976; Aguilar \& White 1985). 
A full treatment of these effects would require a 
dynamical simulation (e.g. Moore et~al. 1996; Gnedin 
2003). Attempts have been made to implement analytic 
descriptions of the phenomena modelled in the simulations into galaxy 
formation models (e.g. Taylor \& Babul 2001; Benson et~al. 2004; 
Yang et~al. 2009). 

In general, standard semi-analytical galaxy formation codes ignore the 
tidal disruption of satellite galaxies. A recent exception  
is the calculation of Henriques, Bertone \& Thomas (2008). These authors 
post-processed the output of the Munich group's semi-analytical model to 
remove galaxies that they believed should have been tidally disrupted. 
Galaxies are associated with the dark matter halo in which they first 
formed as a central galaxy. When this halo merges with a more massive halo, 
it becomes a satellite halo or substructure, and is stripped of mass through 
dynamical effects. Eventually, the substructure may fall below the resolution 
limit of the N-body simulation (in this case the Millennium Simulation). 
Henriques et~al. removed satellites whose host dark matter substructure 
had dissolved, and added these to the ICL. They found that by 
adopting this procedure, the model predictions agreed better with the slope 
of the faint end of the luminosity function and the colour distribution 
of galaxies. However, this algorithm depends on the resolution of the N-body 
simulation, which governs when subhalos are destroyed. Moreover, the softening 
length adopted in the simulation exceeds the scale size of all but the very 
brightest galaxies. Hence, it is not clear that any of the more condensed 
baryonic material would have been stripped from the model galaxies, even 
when the host dark matter halo has been shredded. 
 
Here we adopt a simpler approach which is independent of 
the resolution of the N-body simulation. We assume that the 
degree of disruption of a satellite 
galaxy depends on the ratio of the mass of the main dark matter halo to 
the mass of the satellite halo at infall, $M_{\rm H}/M_{\rm sat}$: 
\begin{equation}\label{ICLapply}
 \frac{L_{\rm new}}{L_{\rm orig}} =  \beta \left( {M_{\rm{H}} \over 
M_{\rm{sat}}}\right)^{-1},
\end{equation}
where $L_{\rm orig}$ is the original luminosity of the satellite galaxy 
predicted by the galaxy formation model, $L_{\rm new}$ is the new 
luminosity intended to take into account stripping of mass from the satellite 
and $\beta$ is an adjustable parameter. We chose this scaling of disrupted 
luminosity fraction because the galaxy merger timescale 
essentially scales with the mass ratio $M_{\rm H}/M_{\rm sat}$; objects 
with large values of  $M_{\rm H}/M_{\rm sat}$ will spend longer orbiting 
within the host dark matter halo and are therefore more susceptible to 
dynamical disruption. Our satellite disruption prescription involves 
post-processing the output of the galaxy formation model, to reduce the 
luminosity of satellite galaxies according to Eq.~\ref{ICLapply}. One 
clear shortcoming of our approach is that we do not take into account 
the time when the satellite galaxy actually fell into the more massive halo. 
With our prescription, a satellite could suffer a large luminosity 
reduction immediately after falling into a larger structure. 
On the other hand, we ignore any stripping which may have occurred at 
earlier stages in the merger history. Hence it is not clear whether 
our simple model for the disruption of satellites is likely to be an 
over or underestimate of the actual effect. 
 
The Millennium Archive does not list the satellite galaxy dark halo mass 
for the Bower et~al. model. Hence, it was necessary for us to re-run the 
Bower et~al. model in order to extract the information required to 
apply the model described by Eq.~\ref{ICLapply}. We present the results 
of rerunning the Bower et~al. model, labelled Bower06-{it bis}, 
{\it without} applying any dynamical disruption, in Fig.~\ref{HODBICL} 
in which we show the HOD and in Fig.~\ref{DFBICL}, where we compare 
the predicted clustering with the 2dFGRS measurements. A 
comparison of the results presented in these plots with the equivalent 
results for the version of the Bower et~al. model available from the 
Millennium Archive (Figs.~\ref{BBN} and ~\ref{DAll} respectively) 
shows a subtle but appreciable change in the model predictions. 
The re-run version of the Bower et~al., which we refer to as 
Bower06-{\it bis}, is actually in better agreement 
with the 2dFGRS clustering results than the Millennium Archive version. 
The main reason for these differences are small improvements in the model. 
There has been substantial code development in the three years since the 
Bower et~al. model was originally placed in the Millennium Archive, 
to incorporate 
new physical ingredients and to improve the implementation of other 
processes. Also, improvements have been made to the construction 
of the dark matter halo merger histories from the Millennium 
(J. Helly, private communication). The re-run Bower et~al. model is 
available in the Millennium archive as Bower06-{\it bis}. 
As we shall see, the changes to the clustering predictions arising 
from the implementation of new physical processes are, in any case, 
larger than those between Bower et~al. and Bower06-{\it bis}. 

The HOD resulting from applying the satellite disruption model of 
Eq.~\ref{ICLapply} is compared with the Bower06-{\it bis} model in 
Fig.~\ref{HODBICL}. The free parameter $\beta$ in the stripping model 
was set to $0.9$ to produce the best match to the clustering measurements, 
as plotted in Fig.~\ref{DFBICL}. As expected, Fig.~\ref{HODBICL} shows 
that there are fewer satellites in the model with disrupted satellites. 
The effect appears largest in the brightest luminosity bin. This is 
primarily due to the imposed change in the shape of the luminosity 
function, rather than to a shift in the typical value of 
$M_{\rm H}/M_{\rm sat}$ for each galaxy sample. 
In the brightest bin, since the 
abundance of galaxies drops exponentially with luminosity, 
more galaxies are shifted out of the bin in the faintwards direction, 
after applying the disruption recipe, than are shifted into that bin 
from brighter luminosities. The change in the HOD generated by  
applying the satellite disruption model  
falls short of the target suggested by the modified HOD derived in the previous section. In the 
intracluster light model, the slope of the satellite HOD is essentially 
unchanged and the biggest variation in the number of satellites is found 
in the brightest luminosity bin rather than the faintest. The resulting 
clustering predictions do not change in the desried way, as shown by 
Fig.~\ref{DFBICL}. Rather than altering the luminosity dependence of 
clustering, the main effect of disrupting satellites is to reduce the 
clustering amplitude in all the luminosity bins. 

We close this section by showing the model prediction for the 
fraction of the total light in a cluster that is in the form of a diffuse 
intergalactic background. Fig.~\ref{LILB} shows that the satellite disruption 
model removes at most 20\% of the total cluster light from galaxies, 
in excellent agreement with the observational estimate from 
Zibetti (2008). This agreement is encouraging as the parameter in 
the satellite disruption model was set without reference to the constraint 
on the background light, but was chosen to 
improve the match to the observed clustering.

\subsection{Mergers between satellite galaxies}
\label{SSmerge}

\begin{figure} 
\includegraphics[width=8.6cm,bb=30 185 570 700]{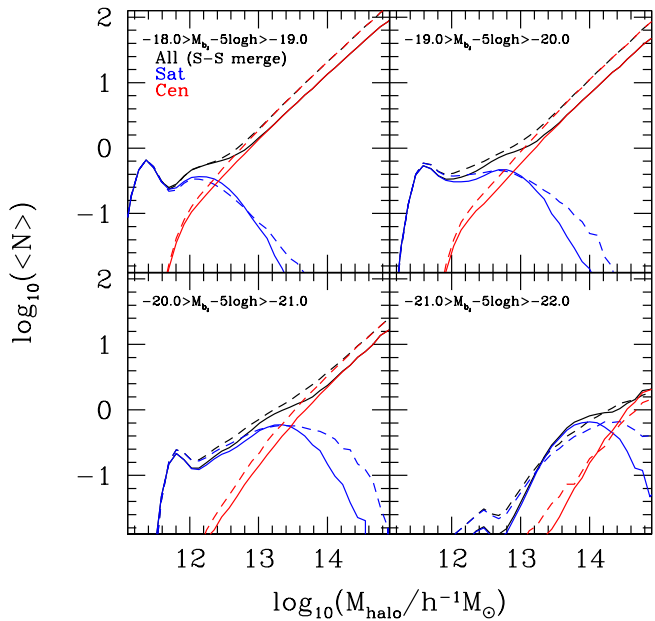}
\caption{\label{HODSS}
The HOD of the model including satellite-satellite mergers (solid lines). 
For reference, the HOD of the Bower06-{\it bis} model is shown by the dashed 
lines. The values of the power-law slope $\alpha$ of the satellite HOD are 
now different in each luminosity bin. 
}
\end{figure}

\begin{figure} 
\includegraphics[width=8.6cm,bb=30 185 570 700]{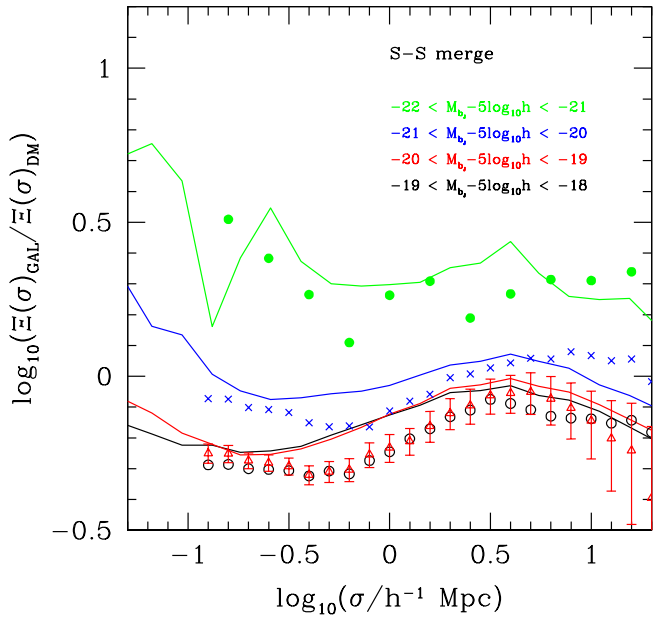}
\caption{\label{DFSS}
The projected correlation functions for galaxies divided by the 
projected correlation function of the dark matter for the 
model with satellite-satellite mergers. The symbols show the 
2dFGRS measurements. The different colours show the different 
luminosity bins. 
}
\end{figure}

Semi-analytical models typically only consider the merger of satellites 
with the central galaxy in a halo. In general, a timescale is calculated 
analytically for the orbit of the satellite to decay due to dynamical 
friction. If this timescale is shorter than the lifetime of the host dark 
matter halo, then the satellite is assumed to merge with the central 
galaxy. When a halo merges with a larger structure, the galaxies in the 
smaller halo are assumed to become satellite galaxies orbiting 
the new central galaxy. 
The satellites retain no memory of the fact that they were 
once members of a common halo. New dynamical friction timescales are 
calculated for each satellite. 

With the advent of ultra-high resolution N-body simulations, there is now 
convincing evidence that this simple picture is incomplete (Springel 
et~al. 2008; Angulo et~al. 2008; Wetzel, Cohn \& White 2009). The simulations 
reveal that, following a merger, the subhaloes of the lower mass halo 
often remain as a distinct unit, orbiting 
coherently in the new main subhalo. Indeed, several levels 
of subhalo hierarchy have been uncovered. By tracing the evolution of the 
subhaloes in these simulations, their ultimate fate can be 
determined. A large fraction of the high mass subhaloes which undergo a 
merger coalesce with the main subhalo of the new halo. However, the 
probability of a merger with a subhalo other than the main subhalo increases 
with decreasing subhalo mass. At $z=0$, Angulo et~al. (2008) found that 
subhaloes with 1\% or less of the total mass of the main subhalo 
were as likely to merge with another subhalo as with the main subhalo. 
Rather than merging with a random subhalo, the merger is with another 
subhalo which shared a common parent halo. A merger which started before 
this parent halo was subsumed by the main halo is being completed inside 
the new halo.  
 
We added satellite-satellite mergers to {\tt GALFORM} by modifying the 
prescription for galaxy mergers. Guided by the results obtained by Angulo 
et~al. for the Millennium Simulation, we modified the calculation of 
the galaxy merger timescale. Depending on the mass ratio, 
$M_{\rm H}/M_{\rm sat}$, and the redshift, we allowed a fraction of satellite 
galaxies to be considered for satellite-satellite mergers (see figure 5 of 
Angulo et~al.). We did this by considering the last but one level of the 
halo merger history i.e. the progenitor haloes of the present day halo. 
For a selected satellite in the progenitor halo, we asked if there 
would be sufficient time for this object to have merged with the central 
galaxy in the progenitor {\it by the present day}, rather than by the end 
of the lifetime of the progenitor. This is equivalent to allowing the merger 
to continue in the substructure after it becomes part of the larger halo. 
If there is sufficient time, then we merge the satellite with the central 
galaxy of the progenitor at the end of the progenitor's lifetime. This means 
that the merger happens sooner than it would do in practice. 
If there is a burst of star formation associated with the merger, 
then this burst will also happen earlier than it should have done. 
However, in the Bower et~al. model there is relatively little star 
formation in bursts at low redshift. Our scheme does, however, reproduce 
the number of satellite-satellite mergers implied by the subhalo mergers 
in the Millennium Simulation. 

By allowing satellite-satellite mergers, we are able qualitatively 
to reproduce the changes suggested by the empirically determined modified 
HOD, as shown in Fig.~\ref{HODSS}. There are two main reasons for the 
change in the HOD. Firstly, satellite-satellite mergers reduce the number 
of satellite galaxies in the model. Secondly, the number of low luminosity 
satellite galaxies in high mass haloes is reduced because these objects 
can merge with other satellites; the remnant is also a satellite but it is, 
of course, brighter than its progenitors.
The HOD for central galaxies also changes, 
with the central galaxies in more massive haloes becoming brighter (and 
hence moving into a brighter luminosity bin). This is because satellites 
which have experienced satellite-satellite mergers are more massive than 
they would have otherwise been and therefore have a shorter dynamical 
friction timescale. The clustering predictions for the model with 
satellite-satellite mergers are shown in Fig.~\ref{DFSS}. The model now 
matches the sequence of luminosity dependent clustering measured in the 
2dFGRS, albeit with slightly higher clustering amplitudes overall.

\subsection{The kitchen sink model}

\begin{figure} 
\includegraphics[width=8.6cm,bb=30 185 570 700]{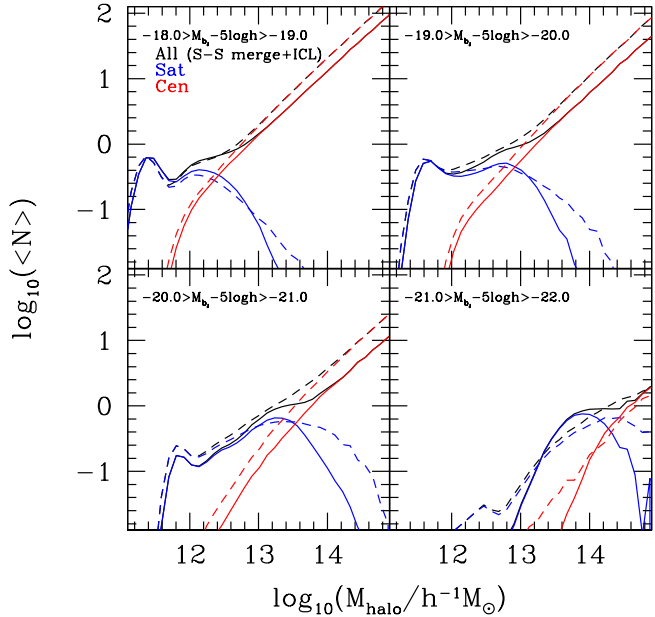}
\caption{\label{HODSSICL} 
The HOD of the hybrid model with satellite-satellite mergers and 
disruption of satellites (solid lines).  The Bower et~al. model HOD is 
shown by the dashed lines. 
}
\end{figure}

\begin{figure} 
\includegraphics[width=8.6cm,bb=30 185 570 700]{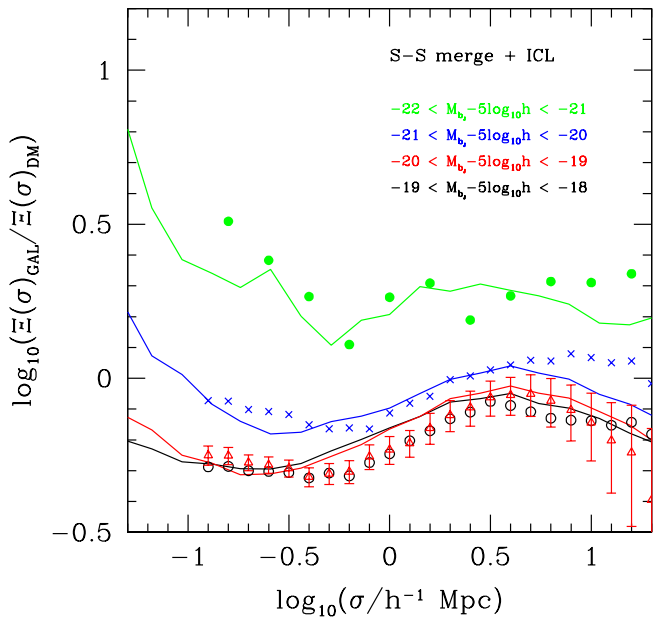}
\caption{\label{DFSSICL} 
The projected correlation functions divided by correlation function 
of the dark matter. The lines show the predictions for the hybrid 
satellite-satellite merger and satellite disruption model. The symbols 
show the 2dFGRS measurements. 
}
\end{figure}

 
\begin{figure} 
\includegraphics[width=8.4cm,bb=30 185 580 700]{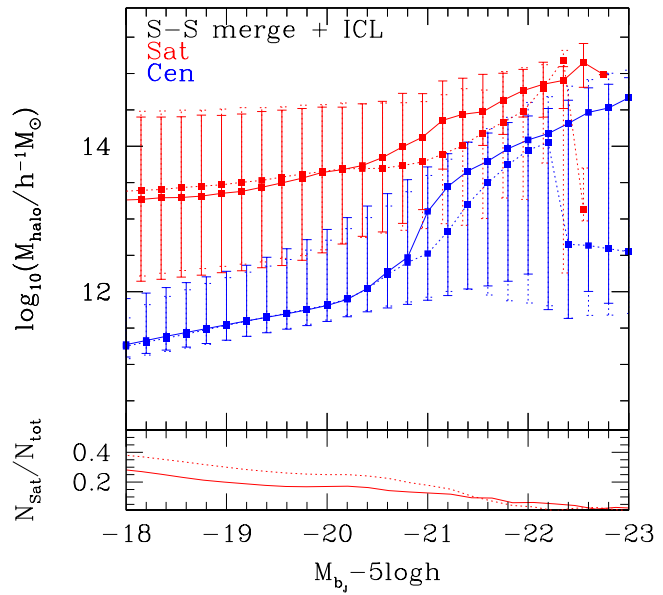}
\caption{\label{MMBJSSICL} 
The host halo mass - luminosity relation for the hybrid model. The upper 
panel shows the median halo mass and the 10-90 percentile range. The red 
points show the relation for satellite galaxies and the blue lines for 
central galaxies. The lower panel shows the fraction of galaxies which 
are satellites as a function of magnitude. The dotted lines in both panels 
show the relations for the original Bower et~al. (2006) model. 
}
\end{figure}

In the previous two subsections we have seen that the satellite disruption 
and satellite-satellite merger models have appealing features. The satellite 
disruption model can change the overall amplitude of the clustering for 
different luminosity samples, whereas the satellite-satellite merger model 
can reproduce the observed trend of clustering strength with luminosity 
if not the precise amplitude. In isolation, neither model offers a fully 
satisfactory solution to the problem of matching the luminosity dependent 
clustering seen in the 2dFGRS. It seems desirable therefore to implement both 
effects in tandem. We do this by generating a model which incorporates  
satellite-satellite mergers and post-processing the resulting 
satellite luminosities using the disruption model of Eq.~\ref{ICLapply}. 

Fig.~\ref{DFSSICL} shows the projected correlation functions predicted by 
the hybrid model. The model predictions are now in remarkably good agreement 
with the 2dFGRS measurements. The model matches the amplitude of clustering, 
the trend and strength of the luminosity dependence of clustering and the 
shape of the correlation functions. The HOD of this model matches the form 
of the reference empirical HOD as shown in Fig.~\ref{HODSSICL}.  
The slope of the satellite HOD in the hybrid model 
is influenced by satellite-satellite mergers, whereas its amplitude 
is determined by satellite disruption. 

Fig.~\ref{LILB} shows how the predicted intracluster light in the hybrid 
model with satellite disruption and satellite-satellite mergers compares 
with the Bower06-{\it bis} model. Again, the amount by which the 
plotted halo luminosity ratio deviates from unity shows the  
fraction of the total light is not attached to galaxies. The 
fraction of intracluster light depends on halo mass and is in very 
good agreement with the observational estimates by Zibetti (2008).

Fig.~\ref{MMBJSSICL} shows the relation between host halo mass and 
galaxy luminosity in the hybrid model. Compared with the Bower et~al. and 
Font et~al. models, there is relatively little difference in the median 
halo mass for either satellite or central galaxies; the changes in 
the median mass are of the order of 0.1dex. However, the host 
halo masses of satellite galaxies are large and thus these haloes are highly 
biased. A small change in the typical host mass will therefore produce 
an appreciable change in the predicted bias. The key difference is in 
the fraction of galaxies that are satellites as a function of magnitude, 
shown in the lower panel of Fig.~15.
The number of satellites in the hybrid model is down by almost a factor 
of two from that in the original Durham models.

\section{Summary and Conclusions}
\label{summary}

The dependence of galaxy clustering on luminosity has been 
measured with high accuracy in the local Universe by the 
2dFGRS and SDSS (Norberg et~al. 2001, 2002; Zehavi et~al. 
2002, 2005; Jing \& Borner 2004; Li et~al. 2006). We have 
shown that the current ``best bet'' publicly available galaxy 
formation models only match the observational results in a 
qualitative sense. These models fail to match the trend 
of clustering strength with luminosity.
We have demonstrated 
that the reason for the discrepancy is that the models predict 
too many satellites in massive haloes. Li et~al. (2007) reached 
a similar conclusion comparing the clustering of galaxies in 
the red selected SDSS with the semi-analytical models of 
Kang et~al. (2005) and Croton et~al. (2006). 

Li et~al. (2007) showed that the match to the observed clustering 
could be improved if $\approx 30\%$ of the satellite galaxies were 
removed from the catalogues generated from the semi-analytical models. 
Li et~al. did this by hand without any reference to the mass of the 
host dark matter halo. This is equivalent to changing the normalization 
of the halo occupation distribution for satellites, without altering  
the slope. In this paper, we first changed the HOD of satellites by 
hand and found that the agreement with the observed clustering could be 
improved by changing the slope of the satellite HOD. For galaxy samples 
close to $L_*$, satellites have to be preferentially removed from more 
massive dark matter haloes. 

Out of the semi-analytical models we considered in this paper, the de Lucia 
\& Blaizot (2007) model came closest to reproducing the observational 
clustering measurements. This was also the model with the smallest number 
of satellites. However, it is not clear to what extent this feature 
is due to approximations 
used in the model (e.g. the adoption of the dynamical time as the 
time over which gas is allowed to cool).
In any case, even this model fails to reproduce the full dependence of 
clustering on luminosity. 

We next tried to remove satellite galaxies from massive haloes 
in the Durham semi-analytical models by perturbing the values 
of the parameters which control certain processes, such as supernova 
feedback, the suppression 
of gas cooling by AGN heating and galaxy mergers. When running a variant 
model, the predicted luminosity function often changes. To ensure that 
changes in the clustering predictions were robust to the requirement that 
a model should reproduce the observed galaxy luminosity function, we rescaled 
the model luminosity functions to agree exactly with the observations. 
We were unable 
to find an improved model within the existing framework, 
which suggests that additional physical processes which mostly affect 
satellite galaxies need to be considered. 

The Durham models have recently been revised as regards the
treatment of gas cooling in satellites (Font et~al. 2008). Satellite 
galaxies can now retain some fraction of the hot halo associated 
with them at infall. The precise fraction depends upon the orbit of 
the satellite. This improvement of the gas cooling treatment alters 
the colours of faint satellites in groups and clusters. The galaxies 
we consider in this paper are brighter by comparison and there is 
little change in the clustering predictions of the Font et~al. model 
compared with those from its predecessor, the Bower et~al. (2006) model. 

In this paper, we considered two processes which are not currently 
included in most galaxy formation models: mergers between satellite 
galaxies and the tidal disruption of satellites. The first of these 
processes is 
motivated by recent high resolution simulations of the formation of 
dark matter haloes which show that hierarchies of substructures persist 
(Diemand et~al. 2008; Springel et~al. 2008). Mergers which started 
in a progenitor halo can run to completion in the descendant halo. The 
disruption of satellites has been modelled analytically 
in the Durham model in a study of the heating of the Milky Way's 
disk (Benson et~al. 2004). Here, we applied a simple prescription to remove 
luminosity from satellites based on the ratio of the host halo mass to 
the mass of the halo in which the satellite formed, which is related 
to the timescale for the satellite's orbit to decay through dynamical 
friction. Applying the model for the disruption of satellites changes 
the overall amplitude of clustering without improving the trend of 
clustering strength with luminosity. 
Including mergers 
between satellites, on the other hand, 
does alter the predictions for the luminosity 
dependence of clustering. By applying both extensions together, we are 
able to obtain a significantly improved match to the 2dFGRS measurements 
(Norberg et~al. 2009). 
The hybrid model matches the observational 
constraints on the amount of intracluster light.

The differences between the clustering predictions of current 
galaxy formation models and observations are small. However, 
the differences can be measured robustly and will become even more 
apparent when larger surveys become available. These discrepancies  
limit the usefulness of the models in the construction of mock 
catalogues needed for the exploitation of future galaxy surveys and 
suggest the need for new physical processes to be incorporated into 
the models. The revisions to the galaxy formation models we propose  
in this paper are simplistic and are merely intended to highlight  
promising areas where the models need to be developed in the future, 
in a self consistent way.

\section*{Acknowledgments}
HSK acknowledges support from the Korean Government's 
Overseas Scholarship. 
CSF acknowledges a Royal Society Wolfson Research Merit Award. 
This work was supported in part by a grant from the Science and Technology 
Facilities Council. We acknowledge helpful conversations with 
Peder Norberg, Enrique Gazta\~{n}aga and Darren Croton; we also 
thank Peder Norberg for supplying data in electronic form in 
advance of publication.

\end{document}